\documentclass[preprint,showpacs,amsmath,amssymb]{revtex4-1}

\usepackage{graphicx}
\usepackage{bm}
\usepackage{color}
\usepackage{amsmath}

\begin{document}

\title{
Sine-square deformation and M\"obius quantization of two-dimensional conformal field theory
}
\author{Kouichi Okunishi}
\affiliation{Department of Physics, Niigata University, Igarashi 2, Niigata 950-2181, Japan}
\date{\today}

\begin{abstract}
Motivated by sine-square deformation (SSD) for quantum critical systems in 1+1-dimension, we discuss a M\"obius quantization approach to the two-dimensional conformal field theory (CFT), which bridges the conventional radial quantization and the dipolar quantization  recently  proposed by Ishibashi and Tada. 
We then find that the continuous Virasoro algebra of the dipolar quantization can be interpreted as a continuum limit of the Virasoro algebra for scaled generators in the SSD limit of the M\"obius quantization approach.
\end{abstract}


\maketitle

\section{introduction}

Recently, intriguing  properties of sine-square deformation (SSD) for 1+1-dimensional quantum many-body systems have been revealed by a series of intensive theoretical researches.\cite{GendiarKN2009,HikiharaN2011,GendiarDLN2011,Katsura2011,MaruyamaKH2011,Okunishi,ShibataH2011,Hotta2,Hotta,Katsura2011B,Tada,Ishibashi1,Ishibashi2} 
The SSD was originally introduced as a spatial deformation of interaction couplings with the sine-square function for 1+1D quantum lattice models.\cite{GendiarKN2009,HikiharaN2011,GendiarDLN2011}
Several numerical studies clarified that critical ground states of SSD systems with the open-boundary are identical to those of the uniform systems with the periodic boundary within numerical accuracy.
Later, this correspondence of the ground states between the uniform and SSD systems was proved for some exactly solved models which can be reduced to free fermionic models.\cite{Katsura2011,MaruyamaKH2011}
In Ref.\cite{Okunishi}, an interesting connection of the SSD to supersymmetric quantum mechanics is pointed out for a nonrelativistic free fermion system in the continuous space.
Moreover, the correspondence of the open and periodic boundary systems under the SSD has been successfully applied to numerical estimation of bulk quantities via finite-size systems\cite{ShibataH2011,Hotta,Hotta2}. 
These scalabilities of the SSD researches suggest that there exists rich physics behind the SSD.

The SSD of the two-dimensional conformal field theory (CFT)\cite{BPZ} was firstly investigated by Katsura,
where the operator $L_0- \frac{L_1+L_{-1}}{2}+ \bar{L}_0- \frac{\bar{L}_1+\bar{L}_{-1}}{2}$ is regarded as a Hamiltonian of the SSD system.\cite{Katsura2011B}
Here, $L_n$ denotes the Virasoro generator of CFT.
Then, an essential point is that the CFT vacuum is annihilated by the deformation terms $\frac{L_1+L_{-1}}{2}$ and $\frac{\bar{L}_1+\bar{L}_{-1}}{2}$, because of SL(2,R) invariance, implying that the ground state of the SSD system is identical to that of the uniform system.
Also, a relevance of the SSD to string theory was discussed on the basis of the free boson in Ref. \cite{Tada}.
Very recently, Ishibashi and Tada has proposed a more direct approach to the SSD of CFT, which is called ``diplolar quantization''; this novel quantization scheme different from the conventional radial quantization provides a continuous Virasoro algebra for the SSD of CFT.\cite{Ishibashi1,Ishibashi2}
However, the connection between the dipolar quantization and the radial quantization, or equivalently between the SSD system and the uniform system, has not been clear at the present stage of researches.
How can we interpolate between the dipolar quantization for the SSD system  and the usual CFT based on the radial quantization?

In order to address the above problem, we introduce the parametarized Hamiltonian that bridges the uniform system and the SSD system\cite{parametarized},
\begin{equation}
{\cal H}= {\cal L}_0 + \bar{\cal L}_0
\label{ssdH_theta}
\end{equation}
with
\begin{equation}
{\cal L}_0\equiv \cosh(2\theta) L_0- \sinh(2\theta) \frac{L_1+L_{-1}}{2}.
\label{ssd_l0_theta}
\end{equation}
where we can assume $\theta \ge 0$ without loss of generality.\cite{Holomorphic}
Note that  $\theta=0$ corresponds to the  uniform system  and $\theta\to \infty$ does to the SSD system except for the overall normalization.
An essential observation of this Hamiltonian is that its ground state is always identical to that of the uniform system of $\theta=0$, since  Eq. (\ref{ssd_l0_theta}) can be regarded as ``Lorentz transformation'' reflecting SL(2,R) symmetry in CFT\cite{Tada}, and $\frac{L_1+L_{-1}}{2}$ annihilates the CFT vacuum.
In this paper,  we derive the classical Virasoro (Witt) algebra based on the Hamiltonian (\ref{ssd_l0_theta}), where the M\"obius transformation coordinate plays a crucial role. 
We then discuss quantization of the Virasoro algebra for (\ref{ssdH_theta}), which we call  ``M\"obius quantization'',  interpolating between the radial quantization and the dipolar quantization. 
In particular, we find that the continuous Virasoro algebra of the dipolar quantization can be interpreted as a continuum limit of the Virasoro algebra obtained in  the SSD limit of the M\"obius quantization.
We also mention that the M\"obius quantization approach gives a consistent result with the M\"obius-type conformal mapping of CFT except at the SSD point.

This paper is organized as follows.
In the next section, we briefly explain the classical Virasoro algebra and the role of the M\"obius coordinate.
In Sec. 3, we explain the M\"obius quantization approach to the central extension of the classical Virasoro algebra.
In particular, we derive the Virasoro algebra for Eq. (\ref{ssdH_theta}) and discuss its continuum limit corresponding to the SSD point.
We also describe the vacuum state and the primary field in the M\"obius quantization.
In Sec. 4, we comment on the conformal mapping associated with the  M\"obius quantization.
In Sec. 5,  we discuss the dipolar/SSD limit of the M\"obius quantization  in details. 
In Appendix A,  the relation of the Virasoro generators between the M\"obius quantization and the radial quantization is provided as a series expansion form, where expansion coefficients are exactly represented by  Gauss's hypergeometric function.

\section{classical Virasoro algebra}

In this section, let us  discuss the classical Virasoro (Witt) algebra for Eq. (\ref{ssd_l0_theta}).
For the usual classical Virasoro algebra, the generator is defined as
\begin{equation}
l_n=-z^{-n+1}\frac{\partial}{\partial z},
\label{classical_l}
\end{equation}
which satisfies the commutation relation
\begin{equation}
[l_n,  l_{n'} ] = (n-n')l_{n+n'} .
\end{equation}

In analogy with Refs.\cite{Ishibashi1,Ishibashi2},  we define a new differential operator for Eq. (\ref{ssd_l0_theta}) as
\begin{equation}
{\mathfrak l}_0\equiv  \cosh(2\theta) l_0 - \sinh( 2\theta) \frac{l_{-1}+l_{1}}{2} =
-g(z) \frac{\partial}{\partial z},
\end{equation}
where
\begin{equation}
g(z)\equiv -\frac{1}{2}\left[\sinh(2\theta)({z}^{2}+1)- 2\cosh(2\theta)z \right] .
\end{equation}
Note that $\theta=0$ describes the uniform system and $\theta \to \infty$ corresponds to the SSD point.
This definition of ${\mathfrak l}_0$ contains a divergent factor if $\theta \to \infty$.
Thus, we also define a scaled generator 
\begin{equation}
\tilde {\mathfrak l}_0\equiv   \frac{{\mathfrak l}_0}{N_\theta},
\label{cl0_n}
\end{equation}
where  $N_\theta$ is the normalization factor,
\begin{equation}
N_\theta \equiv \cosh(2\theta).
\end{equation}
It should be noted that the scaled generator $\tilde{\mathfrak l}_0$ was dealt with  in the previous papers\cite{Ishibashi1,Ishibashi2}, where the dipolar quantization at the SSD point ($\theta \to \infty$) was directly discussed.

According to Refs.\cite{Ishibashi1,Ishibashi2}, we consider the eigenfunction of ${\mathfrak l}_0$,
\begin{equation}
{\mathfrak l}_0 f_\kappa(z) = -\kappa N_\theta  f_\kappa(z),
\end{equation}
where we have factorized out the normalization factor $N_\theta$ for later convenience.
Assuming that this eigenfunction is single valued on the complex plane, we have a discrete eigenvalue
\begin{equation}
\kappa N_\theta = n
\label{cspectrum_n} 
\end{equation}
with an integer index $n$, and the corresponding eigenfuction
\begin{equation}
f_n(z)
= A_n \left(\frac{1- \cosh(2\theta) +  \sinh(2\theta) z}{1+ \cosh(2\theta) -  \sinh(2\theta)z } \right)^{n}
=A_n (-\tanh(\theta) )^{n} \left(\frac{\sinh(\theta) -  \cosh(\theta) z}{\cosh(\theta) -  \sinh(\theta)z } \right)^{n}
\end{equation}
where $A_n$ is a normalization constant.

On the basis of $f_n(z)$, we next define the differential operator with an index $n$,  
\begin{equation}
{\mathfrak l}_n = 
-g(z) f_n(z) \frac{\partial}{\partial z} .
\end{equation}
Then, a straightforward calculation yields 
\begin{equation}
[{\mathfrak l}_n, {\mathfrak l}_{n'}] =(n'-n) 
g(z) f_n(z) f_{n'}(z)\frac{\partial}{\partial z}
\label{tag_fn}
\end{equation}
If $A_n =  (\tanh(\theta))^{- n }$ is adopted,  $A_n A_{n'} =  A_{n+n'}$ is satisfied.
We can then verify $f_n(z) f_{n'}(z) =  f_{n+n'}(z)$ with
\begin{equation}
f_n(z)=(-)^n \left(\frac{\sinh(\theta) -  \cosh(\theta) z}{\cosh(\theta) -  \sinh(\theta)z } \right)^{n},
\label{classical_fn}
\end{equation}
and Eq. (\ref{tag_fn}) becomes 
\begin{equation}
[{\mathfrak l}_n, {\mathfrak l}_{n'}] =(n-n'){\mathfrak l}_{n+n'},
\end{equation}
which is nothing but the classical Virasoro algebra as well.

An interesting point on the eigenfunction $f_n(z)$ is that it contains the identical form to the M\"obius transformation of SL(2,R) for the complex variable $z$.
If $\theta =0 $, $f_n(z) = -z^{n}$, which reproduces the conventional classical Virasoro algebra of the radial quantization approach. 
 In the SSD limit ($\theta\to \infty$), on the other hand, $\kappa = n/N_\theta$ becomes continuous, and
\begin{eqnarray}
f_\kappa(z)=&&\left(\frac{1- \cosh(2\theta) +  \sinh(2\theta) z}{1+ \cosh(2\theta) -  \sinh(2\theta)z } \right)^{\kappa N_\theta}
\nonumber \\
\simeq&& \left(\frac{1 - \frac{1}{1-z}\frac{1}{N_\theta}  }{1+ \frac{1}{1-z} \frac{1}{N_\theta}  } \right)^{\kappa N_\theta}  \xrightarrow{N_\theta\to \infty}\,  \exp\left(\frac{2\kappa}{z-1}\right),
\end{eqnarray}
which is the same as the kernel function obtained by Ishibashi and Tada\cite{Ishibashi1,Ishibashi2}.
Then, a scaled differential operator $\tilde{\mathfrak l}_\kappa\equiv {\mathfrak l}_n/N_\theta $ becomes a generator of the classical Virasoro algebra with the continuous index $\kappa$.

\subsection{M\"obius coordinate}

We briefly analyze the relation of ${\mathfrak l}_0$ with the M\"onius coordinate, which plays an essential role in quantizing  $ {\mathfrak l}_n $.
As in the usual CFT, we may respectively regard a time development operator and a spatial translation operator as ${\mathfrak l}_0+\bar{\mathfrak l}_0$ and ${\mathfrak l}_0-\bar{\mathfrak l}_0$.
On this basis, we can specify the complex coordinate $\zeta=\tau+i s$ with
\begin{eqnarray}
&&-\frac{\partial}{\partial \tau} = {\mathfrak l}_0 + \bar{\mathfrak l}_0 ,
\label{tau_l0} \\
&&-\frac{\partial}{\partial s} = i({\mathfrak l}_0 - \bar{\mathfrak l}_0) ,
\label{s_l0}
\end{eqnarray}
which leads us to
\begin{equation}
\zeta=\tau+is=  \ln \left(-\tanh(\theta) \frac{\sinh( \theta) -\cosh( \theta)z}{\cosh( \theta) -\sinh( \theta)z}  \right)
\label{zetacoordinate}
\end{equation}
Then, the real part of this equation is reduced to 
\begin{equation}
\coth(\theta)e^{\tau}= \left| \frac{\sinh( \theta) -\cosh( \theta)z}{\cosh( \theta) -\sinh( \theta)z}  \right|,
\label{trajec1}
\end{equation}
where we have assumed $\theta\ge 0$ without loss of generality. 
Writing $z=x+iy$, we can draw constant-$\tau$ contours in the $z$ plane.
For $\tau\ne 0$, we have a circle of the radius $R$ and the center $(X,0)$:
\begin{eqnarray}
\left(x -X \right)^2 +y^2 =R^2 ,
\label{trajec2}
\end{eqnarray}
where 
\begin{equation}
X= \frac{1}{e^{2\tau}-1}\frac{1}{\sinh(\theta)\cosh(\theta)}+\frac{1}{\tanh(\theta)},
\label{xcenter}
\end{equation}
 and 
\begin{equation}
R=\frac{e^{\tau}}{|e^{2\tau}-1|}\frac{1}{\sinh(\theta)\cosh(\theta)}.
\label{rradius}
\end{equation}

In Fig. 1, we show a typical time-flow  diagram with constant-$\tau$ contours for $\theta=1$. 
At $\tau=-\infty$, the infinitely small circle is located at $X=\tanh(\theta)$, indicating the source of the time flow.
As $\tau$ increases to $0$, the center position moves toward $X= -\infty$, and, at the same time, the radius $R$ also becomes larger. 
At $\tau=0$, the contour (\ref{trajec1}) reduces to the vertical line of
\begin{equation}
x= 1/\tanh (2\theta),
\end{equation}
with
\begin{equation}
s=2  \arctan(\sinh(2\theta)y),
\label{s_tau_zero}
\end{equation}
which runs over  $[-\pi , \pi ]$.
As $\tau$ develops beyond $\tau=0$,  $X$ moves from $X=\infty$ to $1/\tanh(\theta)$, with the radius $R$ shrinking. 
Thus, the time flow is described by the gradient flow against the constant $\tau$ circles, departing from the source of $(\tanh(\theta), 0)$ at $\tau=-\infty$ and arriving at the sink of $(\tanh(\theta), 0)$ at $\tau=\infty$.

\begin{figure}
\includegraphics[width=0.4\linewidth]{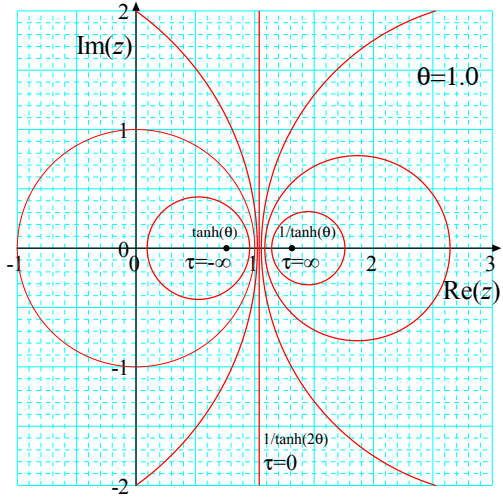}
\caption{The time flow in the M\"obius coordinate and constant-$\tau$ contours for $\theta=1$.}
\end{figure}

We next consider the spatial range of $s$ for a fixed $\tau$.
Taking the imaginary part of Eq. (\ref{zetacoordinate}), we obtain
\begin{equation}
s=  \arctan\left(\frac{(e^{2\tau}-1)\sin\phi}{(e^{2\tau}+1)\cos\phi+2e^{\tau}} \right),
\end{equation}
where we have introduced the parametarization,
\begin{equation}
x=X+R\cos\phi,\qquad y= R\sin\phi
\end{equation}
for a constant-$\tau$ circle.
An important point is that, if $\phi$ runs over  $[-\pi, \pi]$,  $s$ also runs over $[-\pi, \pi]$, independent of $\tau$ and $\theta$\cite{arctan}. 
Together with (\ref{s_tau_zero}) for the $\tau=0$ case, we conclude that the spatial coordinate $s$ always moves in
 \begin{equation}
- \pi  \le s \le  \pi ,
\label{srange}
\end{equation}
which is a finite strip of the width $2 \pi $ independent of $\tau$ and $\theta$.

\section{M\"obius quantization}

In the usual CFT, the stress tensor $T(z)$(and $\bar{T}(\bar{z})$) is formally expanded by the Laurent series around $z=0$ as 
\begin{equation}
 T(z)= \sum_n z^{-n-2} L_n
\end{equation}
which defines the conventional  generator of the  Virasoro algebra. 
In this section, we directly discuss the central extension of the Virasoro algebra on the M\"obius coordinate and its relation to $L_n$.

\subsection{conserved charge and generator}
\label{subsec3-a}

On the basis of the space-time structure of $\zeta$ in the previous section, we consider a conserved charge of the stress tensor for the constant-$\tau$ contour 
\begin{eqnarray}
{\cal L}_n &=&\oint_\tau\frac{dz}{2\pi i}
g(z)f_n(z) T(z) \nonumber \\
&=&
 -(-)^n\frac{\sinh(2\theta)}{2\tanh^n(\theta)} \oint_\tau \frac{dz}{2\pi i}\frac{(z- \tanh(\theta))^{n+1}}{(z-1/\tanh(\theta))^{n-1}} T(z) ,
\label{defln_mobius}
\end{eqnarray}
where the suffix $\tau$ of the integral indicates that the contour integral is performed along the circle of $\tau=$const surrounding the time source of $z=1/\tanh(\theta)$.

The integrand of  Eq. (\ref{defln_mobius}) has no singularity for $n=0,\pm 1$, reflecting the global conformal symmetry of SL(2,R).
Then, setting up the contour surrounding $z=0$ and $1/\tanh(\theta)$,
we  have 
\begin{equation}
{\cal L}_0 = \cosh(2\theta)L_0 - \sinh(2\theta) \frac{L_1+L_{-1}}{2},
\label{ssdl0}
\end{equation}
which reproduces the Lorentz transformation of Eq. (\ref{ssd_l0_theta}).
We can also obtain the closed form of ${\cal L}$ for $n=\pm 1$,
\begin{eqnarray}
{\cal L}_1&=& \frac{\sinh(2\theta)}{2}\left( \frac{1}{\tanh(\theta)}L_1 -2 L_0 + \tanh(\theta)L_{-1}\right)\, , \label{ssdl1} \\
{\cal L}_{-1}&=& \frac{\sinh(2\theta)}{2}\left(\tanh(\theta) L_1 -2 L_0 + \frac{1}{\tanh(\theta)}L_{-1}\right) \, .\label{ssdl-1}
\end{eqnarray}
Moreover, these operators satisfy the sl(2) subalgebra
\begin{equation}
[{\cal L}_1,{\cal L}_{-1}]=2{\cal L}_0, \qquad [{\cal L}_0, {\cal L}_{\pm 1}]=\mp {\cal L}_{\pm 1}.
\end{equation}
For calculation of $n >1 $, the algebraic series expansion with respect to $z \tanh(\theta)$ gives 
\begin{eqnarray}
{\cal L}_n =-(-)^n\frac{\sinh(2
\theta)}{2 (\tanh(\theta))^{-n}}\oint \frac{dz}{2\pi i}(1-z/\tanh(\theta))^{n+1} (\sum_{m=0}^\infty (z\tanh(\theta))^{m} )^{n-1} \sum_l z^{-l-2} L_l
\label{lnplus}
\end{eqnarray}
For $n=2$, for instance, we  explicitly have  
\begin{eqnarray}
{\cal L}_{2} &=&  -\frac{\sinh(2\theta)}{2}\big[ t^2 L_{-1} + \left( {t}^{3}-3\ t \right) L_{{0}}+ \left( {t}^{4}-3{t}^{2}+3  \right) L_{{1}}
+ \left({t}^{5}-3{t}^{3}  +3t -\frac{1}{t} \right) L_{{2}} \nonumber 
\\&& \left. 
+ \left( {t}^{6}-3{t}^{4}+3{t}^{2}-1 \right) L_{{3}}
\cdots  \right]
\label{l2plus}
\end{eqnarray}
where $t\equiv \tanh(\theta)$.
In general,  expansion coefficients in front of $L_l$ can be described by a polynomial with respect to $\tanh(\theta)$, the detail of which is presented in Appendix A.

For the analytic continuation of $n \le -2$, we expand the integrand of (\ref{defln_mobius}) around the $z=\infty$, because $z/\tanh(\theta)$ is beyond the convergence radius of the algebraic series.
Using $w=1/z$, we have 
\begin{eqnarray}
{\cal L}_n =-(-)^n\frac{\sinh(2\theta)}{2\tanh^n(\theta)} \oint \frac{dw}{2\pi i}(1-w/\tanh(\theta))^{-n+1} (\sum_{m=0}^\infty (w\tanh(\theta))^{m} )^{-n-1} \sum_l w^{l-2} L_l .
\label{lnminus}
\end{eqnarray}
For $n=2$, we have
\begin{eqnarray}
{\cal L}_{-2}&=&-\frac{\sinh(2\theta)}{2} \big[ 
{t}^{2}L_{{1}}+
 \left( {t}^{3}-3 t \right) L_{{0}}
+ \left( {t}^{4}-3\,{t}^{2}+3 \right) L_{{-1}}
+\left({t}^{5}-3\,{t}^{3}+3\,t  -\frac{1}{t}\right) L_{{-2}}\nonumber \\
&&+ \left( {t}^{6}-3\,{t}^{4}+3\,{t}^{2}-1\right) L_{{-3}}
+ \cdots \big] .
\label{l2minus}
\end{eqnarray}

Here, we make some comments on the above results.
First, we can easily see ${\cal L}_{\pm 2} \to L_{\pm 2}$ in the $\theta \to 0$ limit, implying that ${\cal L}_{n}$ recovers the conventional Virasoro generator of the radial quantization. 
In the SSD limit($\theta \to \infty$), on the other hand, we can see that all of ${\cal L}_{\pm 1}$ and ${\cal L}_{\pm2}$ reduce to ${\rm const} \times  {\cal L}_0$. 
This suggests that ${\cal L}_n$ of a finite $n$ collapses to ${\cal L}_0$ in the $\theta \to \infty$ limit, reflecting  $\kappa=n/N_\theta \to 0$.
In other words, a continuous spectrum may be well defined for $n,\, N_\theta \to \infty$ with fixing $n/N_\theta=\kappa$, as will be seen  in the next subsection.

Next, we discuss the Hermitian conjugate of the generator ${\cal L}_n$.
We can see that  Eqs. (\ref{lnplus}) and (\ref{lnminus})  have the symmetric form with respect to exchanging $n\leftrightarrow -n$.
If $L_n$ is equipped with the standard Hermitian conjugate of $L_{-n}=L_n^\dagger$,  this symmetric relation of (\ref{lnplus}) and (\ref{lnminus})  ensures the Hermiticity of the generator,
\begin{equation}
{\cal L}_{-n} = {\cal L}_n^\dagger, 
\label{hermite_ssd}
\end{equation}
as long as the series is convergent. 
Comparing (\ref{l2plus}) and (\ref{l2minus}), indeed,  we can easily  verify ${\cal L}_{-2} = {\cal L}_2^\dagger$ for $n=2$.
However, it should be remarked that the convergence radii of the geometric series used in Eqs. (\ref{lnplus}) and (\ref{lnminus}) require $|\tanh(\theta)|<1$.
Thus, the conventional Hermitian conjugate of Eq. (\ref{hermite_ssd}) may not be the case just at the SSD point where $\tanh(\theta)=1$.
As was pointed out in Ref. \cite{Ishibashi2}, another definition of the Hermitian conjugate may be needed at the SSD point.

\subsection{Virasoro algebra}

We next construct the commutator for the generator ${\cal L}_n$.
The operator product expansion (OPE) of the stress tensor is given by
\begin{equation}
T(z)T(w)\sim \frac{c/2}{(z-w)^4}+\frac{2T(w)}{(z-w)^2}+\frac{\partial_w T(w)}{(z-w)}\cdots.
\end{equation}
where $c$ is the central charge.
Then, we can directly calculate the commutator with the contour integrals following the time ordering in the M\"obius quantization:
\begin{eqnarray}
&&[{\cal L}_m, {\cal L}_n]\nonumber\\
&=&  
 \oint_{C_>}\frac{dz}{2\pi i}g(z)f_m(z) \oint_{C}\frac{dw}{2\pi i} g(w)f_n(w) T(z)T(w)  - 
 \oint_{C}\frac{dw}{2\pi i} g(w)f_m(w)
 \oint_{C_<}\frac{dz}{2\pi i} g(z)f_n(z) T(w)T(z)   \nonumber \\
&=& 
\oint_{C}\frac{dw}{2\pi i} g(w) f_m(w)
 \oint_{C_w}\frac{dz}{2\pi i} g(z) f_n(z)  \left[ \frac{c/2}{(z-w)^4}+\frac{2T(w)}{(z-w)^2}+\frac{\partial_w T(w)}{(z-w)}\cdots \right], \nonumber 
\end{eqnarray}
where the integration contours are depicted in Fig. 2.
Substituting $g(z)$ and $f_n(z)$ in the above expression and performing the contour integral of $z$, we obtain
\begin{eqnarray}
&&[{\cal L}_m, {\cal L}_n]\nonumber\\
&=& (-)^{n+m}\frac{\sinh^2(2\theta)}{4\tanh^{m+n}(\theta)}\int_{C}\frac{dw}{2\pi i} 
\left\{ \frac{cn(n-1)(n+1)}{12(\sinh(\theta)\cosh(\theta))^3}\frac{(w-\tanh(\theta))^{n+m-1}}{(w-\frac{1}{\tanh(\theta)})^{n+m+1}}\right. \nonumber\\
&&\left. +   \left[ 2 \left(\frac{(w-\tanh(\theta))^{m+1}}{(w-\frac{1}{\tanh(\theta)})^{m-1}}\right)\left(\frac{(w-\tanh(\theta))^{n+1}}{(w-\frac{1}{\tanh(\theta)})^{n-1} }\right)'
 - \left(\frac{(w-\tanh(\theta))^{m+1}}{(w-\frac{1}{\tanh(\theta)})^{m-1}}\frac{(w-\tanh(\theta))^{n+1}}{(w-\frac{1}{\tanh(\theta)})^{n-1} }\right)' \right] T(w)\right\},
\nonumber 
\end{eqnarray}
where we have used the integration by part for the last term in the square bracket.
Then, a straightforward calculation of the $w$ integral yields the Virasoro algebra of the generator ${\cal L} $ of the M\"obius quantization:
\begin{equation}
[{\cal L}_m, {\cal L}_n]=(m-n){\cal L}_{m+n}+   \frac{c}{12}m(m-1)(m+1)\delta_{n+m,0}  \, .
\label{dvirasoro}
\end{equation}
Thus, the Lorentz transformation (\ref{ssd_l0_theta}) basically provides the same physics as CFT based on the radial quantization. 
Moreover, this fact indicates that the M\"obius quantization can be obtained through the conformal map of the M\"obius transformation. We will comment this point in the next section.
Nevertheless, we would like to emphasize that the M\"obius quantization clarifies an origin of the continuous Virasoro algebra obtained by the dipolar quantization, as follows.

\begin{figure}[t,b]
\includegraphics[width=0.7\linewidth]{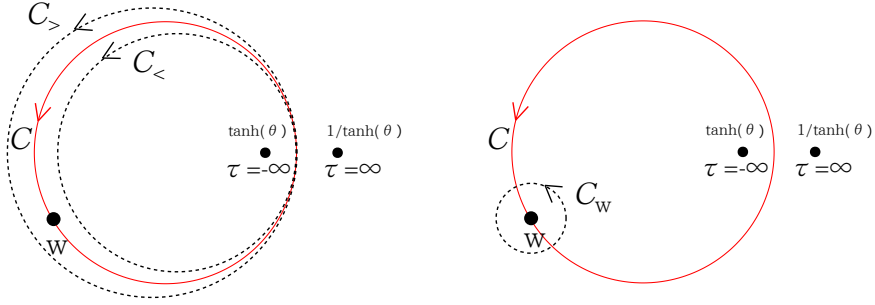}
\caption{The contours in the left panel can be deformed to those in  the right panel.}
\end{figure}

An important viewpoint of Eq. (\ref{dvirasoro}) is  that the SSD limit, i.e.  $N_\theta(=\cosh(2\theta))\to \infty$ should correspond to the dipolar quantization by Ishibashi and Tada.
Recalling Eqs. (\ref{cl0_n}) and (\ref{cspectrum_n}), we define the scaled spectrum and generator as 
\begin{equation}
n/N_\theta \to \kappa , \quad  {\cal L}_n/N_\theta \to \tilde{\cal L}_\kappa ,\quad  N_\theta \delta_{n,0} \to  \delta(\kappa).
\label{def_scaled_ln}
\end{equation}
In the $N_\theta\to \infty$ limit, then, $\kappa$ becomes the continuous index and Eq. (\ref{dvirasoro}) is reduced to
\begin{equation}
[\tilde{\cal L}_\kappa, \tilde{\cal L}_{\kappa'}]=(\kappa-\kappa')\tilde{\cal L}_{\kappa+\kappa'}+   \frac{c}{12}\kappa^3\delta(\kappa+\kappa') ,
\label{cvirasoro}
\end{equation}
which is so called continuous Virasoro algebra.
In this sense, the continuous Virasoro algebra of the dipolar quantization can be interpreted as a continuum limit of the Virasoro algebra for the scaled generator.

\subsection{Vacuum}
In the usual CFT based on the radial quantization, the vacuum is defined as 
\begin{equation}
|0\rangle = I(z=0) |0\rangle\quad {\rm and}\quad   L_n |0\rangle = 0, \quad {\rm for }\quad  n\ge -1
\end{equation}
where $I(z)$ is the identity operator.
Then, the Hermiticity $L_n^\dagger = L_{-n}$ ensures 
\begin{equation}
\langle 0| L_n  = 0, \quad {\rm for }\quad  n\le 1
\end{equation}

For the case of M\"obius quantization, we may define the vacuum as 
\begin{equation}
|0\rangle_{\theta} = I(z=\tanh(\theta)) |0\rangle_{\theta} \quad {\rm and}\quad    {\cal L}_n |0\rangle_{\theta} = 0, \quad {\rm for }\quad  n\ge -1,
\end{equation}
where $|0\rangle_{\theta}$ denotes the vacuum for the M\"obius quantization.
The Hermiticity of Eq. (\ref{hermite_ssd}) gives
\begin{equation}
_{\theta}\langle 0| {\cal L}_n  = 0, \quad {\rm for }\quad  n\le 1
\end{equation}

Here, we should recall that  Eqs. (\ref{ssdl0}), (\ref{ssdl1}), (\ref{ssdl-1}) and (\ref{lnplus}) imply that ${\cal L}_n$ for $n\ge -1$ can be expanded by only  $L_n$ with $n\ge -1$.
Thus, we trivially have
\begin{equation}
|0\rangle_{\theta} =|0 \rangle,
\end{equation}
which is also a consequence of the SL(2,R) invariance of the SSD Hamiltonian.

\subsection{Primary field}

We next discuss the primary field.
In general, for a conformal transformation $z\to w(z)$, a primary field of a scaling dimension $h$ is defined by
\begin{equation}
\phi'(w)=\left(\frac{d z}{d w} \right)^{h} \phi(z),
\end{equation}
which leads us to the commutator
\begin{equation}
[L_n, \phi_h(z)]=z^{n+1}\partial_z \phi_h(z) + (n+1)h z^n \phi_h(z),
\end{equation}
or, equivalently, the OPE 
\begin{equation}
T(z)\phi_h(w) = \frac{h\phi_h(w)}{(z-w)^2} -\frac{\partial_w \phi_h(w)}{z-w}\cdots.
\label{ope_primary}
\end{equation}
In particular,  the primary field specifies the highest weight state  as  
\begin{equation}
L_0 \phi_h(x) = h \phi_h(x) \quad {\rm or }\quad L_0 |h\rangle = h|h\rangle,
\end{equation}
where $|h\rangle \equiv \phi_h(0)|0\rangle$.

For the M\"obius quantization, we have the commutator
\begin{eqnarray}
[{\cal L}_n, \phi_h(z)]&=& \oint_z\frac{dw}{2\pi i}g(w)f_n(w) T(w) \phi_h(z)\nonumber \\
&=&-(-)^n\frac{\sinh(2\theta)}{2(\tanh(\theta))^n}\left[ h  \left(\frac{(z-\tanh(\theta))^{n+1}}{(z-\frac{1}{\tanh(\theta)})^{n-1}}\right)' \phi_h(z) -\left(\frac{(z-\tanh(\theta))^{n+1}}{(z-\frac{1}{\tanh(\theta)})^{n-1}} \right)\partial_z \phi_h(z) \right],
\label{stress_primary}
\end{eqnarray}
where we have used the OPE (\ref{ope_primary}).
Particularly for $n=0$, we have explicitly
\begin{eqnarray}
[{\cal L}_0, \phi_h(z)]= -\frac{\sinh(2\theta)}{2} &&\left[ h\left((z-\tanh(\theta))+(z-\frac{1}{\tanh(\theta)})\right)\phi_h(z) \right.\nonumber \\
&&  -\left. (z-\tanh(\theta))(z-\frac{1}{\tanh(\theta)})\partial_z\phi_h(z)
 \right]
\end{eqnarray}
If we define the primary state for the M\"obius quantization as 
\begin{equation}
|h\rangle_{\theta} = \phi_h(z=\tanh(\theta))|0\rangle,
\label{primary_m}
\end{equation}
we can easily see 
\begin{equation}
{\cal L}_0 |h\rangle_{\theta}=  h |h\rangle_{\theta}.
\label{ssdl0_primary}
\end{equation}
From Eq. (\ref{stress_primary}), moreover,  we straightforwardly read,
\begin{equation}
{\cal L}_{n} |h\rangle_{\theta}= 0, \quad {\rm for} \quad n\ge 1 .
\end{equation}
For $n\le -1$, on the other hand, ${\cal L}_n$ generates a new state ${\cal L}_n|h\rangle_{\theta}$, as in the usual CFT.
Using the commutator (\ref{dvirasoro}) for ${\cal L}_n$,  we can thus construct the representation of the Virasoro algebra for the primary state (\ref{primary_m}).


The relation between $|h\rangle_{\theta}$ and $|h\rangle$ can be directly seen through the conventional Virasoro generator by the radial quantization; 
Since $L_{-1}$ is the translation operator in the $z$ plane, we have 
\begin{equation}
\phi_h(\alpha)=e^{\alpha L_{-1}}\phi_h(0)e^{-\alpha L_{-1}}.
\end{equation}
Then, Eq. (\ref{primary_m}) is rewritten as 
\begin{equation}
|h\rangle_{\theta} = e^{\alpha L_{-1}}\phi_h(0)e^{-\alpha L_{-1}}|0\rangle = e^{\alpha L_{-1}}\phi_h(0)|0\rangle =e^{\alpha L_{-1}}|h\rangle
\label{primary_rm}
\end{equation}
with $\alpha=\tanh(\theta)$.
Using the commutators in Appendix B, we can calculate
\begin{eqnarray}
&&  \left(\cosh(2\theta)L_0-\frac{\sinh(2\theta)}{2}(L_{-1}+L_{1})\right)e^{\alpha L_{-1}}|h\rangle \nonumber\\
&&= \left[  h\left(\cosh(2\theta)- \alpha\sinh(2\theta)  \right)
+ \left(\alpha \cosh(2\theta)-\frac{\sinh(2\theta)}{2}(\alpha^2+1) \right) L_{-1}\right] e^{\alpha L_{-1}}|h\rangle,
\end{eqnarray}
which enables us to directly verify  
\begin{eqnarray}
 {\cal L}_0 e^{\tanh(\theta) L_{-1}} |h\rangle = h  e^{\tanh(\theta) L_{-1}}|h\rangle.
\end{eqnarray}
This result is of course consistent with Eq. (\ref{ssdl0_primary}).
Moreover, the norm of $|h\rangle_{\theta}$ can be evaluated as
\begin{eqnarray}
_{\theta}\langle h|h \rangle_{\theta} =\langle h|e^{\alpha L_1}e^{\alpha L_{-1}}|h \rangle=1+\sum_{n=1}^\infty B_n ,
\label{hm_norm}
\end{eqnarray}
with
\begin{equation}
B_n=\frac{\alpha^{2n}}{n!}(n-1+2h)(n-2+2h)\cdots(1+2h)(2h) .
\end{equation}
If $\alpha=\tanh(\theta) <1$, 
\begin{equation}
\left|\frac{B_{n+1}}{B_n}\right|=\alpha^2(1+\frac{2h-1}{n+1}) < 1 \qquad {\rm for}\quad n > \, ^\exists n_c \, .
\end{equation}
Thus, the series of Eq. (\ref{hm_norm}) is convergent for $\tanh(\theta) <1 $, implying that $|h\rangle_{\theta}$ is normalizable.
In the SSD limit, however, the series is divergent\cite{Tada}. 
This suggests that it may be difficult to construct the primary field just at the SSD point within the present approach where the series of (\ref{lnplus}) and (\ref{lnminus}) are also divergent.
In this sense, the primary field for the diplolar quantization is a nontrivial problem.

\section{The conformal mapping approach}

In the previous section, we have quantized  CFT on the M\"obius coordinate, which yields the Virasoro algebra of the same form as that of the radial quantization.
Of course, this is a natural consequence of the SL(2,R) symmetry in CFT, which suggests a more direct approach to obtain the Virasoro algebra of ${\cal L}$.
Let us consider the M\"obius-type conformal map of SL(2,R).
\begin{equation}
w= -\frac{\sinh(\theta)-\cosh(\theta)z}{\cosh(\theta)-\sinh(\theta) z}
\label{mobiusmap}
\end{equation}
We then define the Virasoro generator in the mapped coordinate($w$-plane) as 
\begin{equation} 
{\cal L}_n \equiv \int\frac{dw}{2\pi i} w^{n+1} T(w).
\end{equation}
Now, we rewrite ${\cal L}_n$ in terms of the $z$-plane.
Recall that, for the conformal map (\ref{mobiusmap}), the stress tensor transforms as 
\begin{equation}
T(w) = \left(\frac{d z}{d w}\right)^2 T(z) =(\cosh(\theta)-\sinh(\theta) z)^4\,T(z) .
\end{equation}
Thus, we have 
\begin{equation}
{\cal L}_n  = \int\frac{dz}{2\pi i}\frac{d w}{d z} w^{n+1} \left(\frac{d z}{d w}\right)^2 T(z) =- (-)^n\frac{\sinh(2\theta)}{2(\tanh(\theta))^n }\int\frac{dz}{2\pi i}\frac{(z-\tanh(\theta))^{n+1}}{( z-1/\tanh(\theta))^{n-1}}  T(z),
\end{equation} 
which is consistent with Eq. (\ref{defln_mobius}).
We can therefore obtain the results of the M\"obius quantization, without passing through complicated analysis of the M\"obius coordinate.
However, the M\"obius quantization approach interpolating between the radial and dipolar quantizations was essential to reveal the continuum limit of the Virasoro algebra.

\section{discussion}

We have discussed the two-dimensional CFT in terms of the M\"obius quantization, which bridges the radial quantization and the dipolar quantization.
An essential feature of the M\"obius coordinate is that the source and sink of the time flow are respectively located at $z=\tanh(\theta)$ and $1/\tanh(\theta)$, and they move toward $z=1$, as the deformation parameter $\theta$ increases.
As far as  $\theta$ is finite, then, we have constructed the Virasoro algebra for the generator ${\cal L}_n$ defined by Eq. (\ref{defln_mobius}),  consistent with the result of the conformal mapping of SL(2,R).
In the SSD ($\theta\to\infty$) limit, the locations of the source and sink finally fuse with each other at $z=1$, which corresponds to the dipolar quantization, where the continuous Virasoro algebra emerges for the scaled generator $\tilde{\cal L}_n$ defined by Eq. (\ref{def_scaled_ln}).

An important implication of the M\"obius quantization approach is that the continuous Virasoro algebra for $\theta\to\infty$ can be viewed as a continuum limit of the conventional Virasoro algebra.
As in the previous works, it is natural to deal with the scaled Hamiltonian $\tilde{H}=\tilde{\cal L}_0 + {\bar{ \tilde{\cal L}}}_0 $ at the SSD point, so that ${\mathfrak l}_0$ and $\bar{\mathfrak l}_0$ in Eqs. (\ref{tau_l0}) and (\ref{s_l0}) should be replaced by  $\tilde{\mathfrak l}_0$ and $\bar{\tilde{\mathfrak l}}_0$.
Then, the $\tilde{\zeta}$ coordinate involves the scale factor $N_\theta$:
\begin{equation}
\tilde{\zeta}=\tilde{\tau}+i\tilde{s} = N_\theta \ln\left(-\tanh(\theta)\frac{\sinh(\theta)-\cosh(\theta)z}{\cosh(\theta)-\sinh(\theta) z}\right).
\label{scaled_zeta}
\end{equation}
which implies that the scale of the spatial coordinate is also modified by $N_\theta$, and the range of $\tilde{s}$ becomes
\begin{equation}
 -\pi N_\theta \le \tilde{s} \le \pi N_\theta,
\end{equation}
in contrast to (\ref{srange}).
Thus, the scaled Virasoro generator $\tilde{\cal L}_n$ has the $\theta$-dependent circumference, $2\pi N_\theta$, which diverges as $\theta\to \infty$.
This is the infinite circumference limit of the dipolar quantization in Refs. \cite{Ishibashi1, Ishibashi2}.

We can also discuss the role of the scale factor $N_\theta$ in analogy with the finite-size scaling of CFT. 
For the radial quantization,  the finite-size-scaling analysis for a cylinder is obtained through the conformal mapping $z'= \frac{\ell}{2\pi} \ln z$, where $\ell$ is a free parameter representing the strip width of the cylinder\cite{cardy}.
For the scaled M\"obius coordinate (\ref{scaled_zeta}), we can rewrite $\tilde{\zeta}$ as 
\begin{equation}
\tilde{\zeta}= \frac{\ell_\theta}{2\pi}\ln w + C,
\label{fss_w}
\end{equation}
with $w$ of Eq. (\ref{mobiusmap}) and $\ell_\theta\equiv 2\pi N_\theta$.
Since $C \simeq -1\,$ for $\theta\gg 1$, this relation can be viewed as a finite-size scaling for the $w$ coordinate.
This suggests that,  as $\ell_\theta$ increases,  the region of the energy scale exhibiting the conformal-tower spectrum for the scaled Hamiltonian $\tilde{H}=\tilde{\cal L}_0 + {\bar{ \tilde{\cal L}}}_0 $ is expected to collapse into the infinitely small energy scale above the ground state.
The numerical finite-size-scaling analysis of the free fermion system at the SSD point actually observed the $1/L^2$ size dependence of the excitation spectrum\cite{Hotta}, instead of the universal $1/L$-dependent spectrum of CFT.
Moreover, the analytic calculation with help of SUSY quantum mechanics for the non-relativistic free fermion provides the exact $1/L^2$ dependence of the spectrum.\cite{Okunishi}
By contrast,  the low-energy spectrum of the unscaled Hamiltonian ${\cal H}={\cal L}_0 +  \bar{\cal L}_0$ may maintain the usual conformal spectrum, where the scale of ${\cal H}$ is magnified by the huge scale factor $N_\theta$.

As was seen above,  some  properties of the spectrum of the scaled generator $\tilde{\cal L}$ can be understand in the analogy with the finite-size scaling.
Also, the continuous Virasoro algebra (\ref{cvirasoro}) for the scaled generator $\tilde{\cal L}$ is basically consistent with the dipolar quantization.
However, we should  mention that there remain some difficulties in the SSD limit.
 $\ell_\theta$ in Eq. (\ref{fss_w}) is not a free parameter but the scale factor embedded in the Lorentz transformation, and thus $\tilde{\zeta}$ has the limit, $\tilde{\zeta}= \frac{2}{z-1}$ for $\theta \to \infty$.
This function corresponds to the mapping of $u=e^{\tilde{\zeta}}=\exp(\frac{2}{z-1})$ containing the essential singularity at $z=1$, which causes peculiar behaviors in the dipolar quantization.\cite{Tada, Ishibashi1, Ishibashi2}.
Accordingly,  the primary field (\ref{primary_m}) is unnormalizable at the SSD point, and the Hermitian conjugate (\ref{hermite_ssd}) is inconsistent with that defined in Ref. \cite{Ishibashi2}.
A significant point of the SSD limit is that the source and sink of the time flow fuse at $z=1$, which never occurs for the case of the usual CFT where $\tau=-\infty$ is assumed to be isolated in the complex plane.
Further investigations are clearly required for understanding the SSD and the continuous Virasoro algebra.
We then believe that the relation (\ref{def_scaled_ln}) provides a solid footing to address the SSD limit.
In addition, it should be noted that a connection to similar quantization approaches based on the global conformal symmetry\cite{Luscher}, such as NS quantization\cite{Rychkov}, is also interesting problem.

\section*{Acknowledgment}

The author would like to thank H. Katsura and T. Tada for useful discussions and comments.
This work was supported in part by Grants-in-Aid No. 26400387  from the Ministry of Education, Culture, Sports, Science and Technology of Japan. 

\appendix

\section{expansion coefficients of ${\cal L}_n$}

The Viraosor generator ${\cal L}_n$ of the M\"obius quantization can be represented by a series of the conventional Virasoro generator $L_l$ of the radial quantization.  We  assume $n>1$ for Eq. (\ref{defln_mobius}).
Using $t=\tanh(\theta)(<1)$ for simplicity, we have 
\begin{eqnarray}
{\cal L}_n & = &
 -(-)^n\frac{\sinh(2\theta)}{2t^n} \sum_l L_l \oint_\tau \frac{dz}{2\pi i}\frac{(z- t)^{n+1}}{(z-1/t)^{n-1}} z^{-l-2} \nonumber \\
& =& -(-)^n\frac{\sinh(2\theta)}{2} \sum_l C^n_l(t) L_l, 
\end{eqnarray}
where the expansion coefficient is defined by
\begin{equation}
C_l^n(t) \equiv  t^{n-l-1} \oint_{(0,1)} \frac{dy}{2\pi i}y^{-l-2} (1-y)^{n+1}(1-t^2 y )^{-n+1} .
\label{defCnl}
\end{equation}
The integration path is the contour surrounding $y=0$ and 1 in the complex $y$ plane.
The pole in the contour is located only at $y=0$, if $n\ge -1$. 
Using Goursat's formula, then,  we have 
\begin{equation}
C_l^n(t) =   t^{n-l-1}\frac{1}{(l+1)!}\left. \frac{d^{l+1}}{d y^{l+1}} (1-y)^{n+1}(1-t^2 y )^{-n+1}\right|_{y=0},
\label{Crodorigues}
\end{equation}
which is a polynomial of $t$. For $n>1$, moreover,  we can represent $C_l^n(t)$ with use of Gauss's hypergeometric function:
\begin{equation}
C_l^n(t)= 
\begin{cases}
(-)^{l+1} \frac{(n+1)!}{(l+1)!(n-l)!}F(n-1,-l-1; n-l+1;t^2)\, t^{n-l-1} &n\ge l \ge -1\\
(-)^{n+1} \frac{(l-2)!}{(n-2)!(l-n)!}F(-n-1,l-1; l-n+1;t^2)\, t^{l-n-1} &n < l 
\end{cases}\, ,
\end{equation}
which is equivalent to Eq. (\ref{Crodorigues}).
If $n=2$, the above formula reproduces Eq. (\ref{l2plus}).
For the case of $n<-1$, ${\cal L}_n$ can be attributed to Eq. (\ref{defCnl}) with $n\rightarrow -n$ after changing the variable $w=1/z$, as in Eq. (\ref{lnminus}).

\section{useful commutators}
For the Virasoro generators $L_0$ and $L_{\pm 1}$, we have the following commutators,
\begin{eqnarray}
[ L_0, (L_{-1})^n ]& =& n(L_{-1})^{n},   \\\
[ L_0, e^{aL_{-1}}] & =& aL_{-1}e^{aL_{-1}}, \\\  
[L_{1}, (L_{-1})^n]&=& n(n-1)(L_{-1})^{n-1} +2n (L_{-1})^{n-1}L_0 , \\\
[L_1, e^{aL_{-1}}]  &=& a^2L_{-1} e^{aL_{-1}}+ 2a e^{aL_{-1}}L_0  ,
\end{eqnarray}
which are useful for calculations in Sec. III


\begin{thebibliography}{99}
\bibitem{Holomorphic} Hereafter,  the antiholomorphic part of generators will not be  shown explicitly, except for some special cases. However we should always suppose the antiholomorphic part parallel to the holomorphic part.
\bibitem{arctan} Thus, $s$ is not the principal value of Arctan.
\end{thebibliography}
\end{document}